\documentclass[structabstract]{aa}  
\usepackage{graphicx}
\usepackage{txfonts}
\usepackage{natbib}
\usepackage{epsfig}
\usepackage{soul}
\begin{document}
   \title{Identification of strong photometric activity in the components of LHS~1070}


   \author{Leonardo A. Almeida \and
          Francisco Jablonski \and
          Eder Martioli
          }

   \institute{Instituto Nacional de Pesquisas Espaciais/MCT \\ 
Avenida dos Astronautas 1758, S\~ao Jos\'e dos Campos, SP, 12227-010, Brazil\\
\email{leonardo@das.inpe.br}
          }


 
  \abstract
   {Activity in low-mass stars is an important ingredient in the evolution of such objects. Fundamental physical properties such as age, 
    rotation, magnetic field are correlated with activity.}
   {We show that two components of the low-mass triple system LHS~1070 exhibit strong flaring activity. We identify the flaring components and obtained an improved astrometric solution for the LHS~1070\,$\mathcal{A}$/($\mathcal{B}$+$\mathcal{C}$) system.}
   {Time-series CCD observations were used to monitor LHS~1070 in the
    B and I$_C$ bands. H-band data were used to obtain accurate astrometry for the LHS~1070\,$\mathcal{A}$/($\mathcal{B}$+$\mathcal{C}$) system.}
   {We have found that two components of the triple system LHS~1070 exhibit photometric activity. We identified that components $\mathcal{A}$ and $\mathcal{B}$ are the flaring objects. We estimate the total energy, $\sim 2.0\times10^{33}$ ergs, and the magnetic field strength, $\sim$ 5.5 kG, of the flare observed in LHS~1070~$\mathcal{B}$. This event is the largest amplitude, $\Delta \rm B \gtrsim 8.2$ mag, ever observed in a flare star.}


\maketitle
%


\section{Introduction}

LHS~1070 is a triple system of low-mass stars at a distance of 
$7.72\pm0.15\,{\rm pc}$ \citep{2005AJ....130..337C}. Component $\mathcal{A}$ has spectral type M5.5-6 and components $\mathcal{B}$ and $\mathcal{C}$ have spectral types M8.5 and M9-9.5, respectively \citep{2000A&A...353..691L}. There is a great deal of interest in this system since the astrometric orbits (specially of components $\mathcal{B}$+$\mathcal{C}$) are well determined, making possible the derivation of precise masses. \citet{2008A&A...484..429S} obtained values of $0.157\pm0.009\,{\rm M}_{\odot}$ for the combined dynamical mass of the $\mathcal{B}$+$\mathcal{C}$ components and $0.272\pm0.017\,{\rm M}_{\odot}$ for the combined dynamical mass of the whole 
system. The masses are close to the H-burning limit making these 
objects interesting targets for detailed studies of the transition between low-mass stars and brown dwarfs.

Besides the low masses, the stars in LHS~1070 present other interesting features: the rotational velocities are $v\sin i\simeq 8, 16, 16\,{\rm km\,s}^{-1}$ for components  $\mathcal{A}$, $\mathcal{B}$ and  $\mathcal{C}$, respectively. While components  $\mathcal{A}$ and  $\mathcal{B}$ present signs of activity with H$\alpha$ emission, component  $\mathcal{C}$ does not \citep{2007A&A...471L...5R}. LHS~1070 also shows intense radio emission \citep{2006ApJ...648..629B}. Under the reasonable assumptions of coevality and spin alignment, \citet{2007A&A...471L...5R} use a Skumanich-like law \citep{1972ApJ...171..565S} with a variable breaking-law index to estimate an age of $\sim 1\,{\rm Gyr}$ for the system. 

In this work we report on photometric evidences for strong activity in the $\mathcal{A}$ and $\mathcal{B}$ components of LHS~1070 and discuss the importance of these results in the context of low-mass stars.


\section{Observations}

The data were collected along an observational program on activity of low-mass stars that is being carried out with the facilities of Laborat\'orio Nacional de Astrof\'{\i}sica (LNA/MCT), in Brazil. We have selected 30 objects in the Southern Hemisphere for at least three differential photometry observing sessions. They were chosen according criteria of being low-mass objects, having suitable comparison stars in a small field-of-view, and being bright enough to be observed even at the 0.6-m telescopes. Table \ref{table:1} summarizes the characteristics
of the data collected for LHS~1070. $N$ is the number of individual images obtained with integration time t$_{\rm exp}$. The H-band images were obtained mainly to improve the astrometric solution for the orbital elements of the pair $\mathcal{A}$/($\mathcal{B}$+$\mathcal{C}$).

\begin{table}
\caption{Log of the photometric observations}             
\label{table:1}      
\centering                          
\begin{tabular}{r c c c c}        
\hline\hline                 
Date~~~~~ & $N$ & \ t$_{\rm exp}$(s) & Telescope & Filter \\    
\hline                        
   Jul 04, 2008 & 140 & 30 & 1.6-m & B \\      
   Jul 05, 2008 & 150 & 30 & 1.6-m & B \\
   Aug 25, 2008 & 653 & 20 & 0.6-m & I$_{\rm C}$ \\
   Aug 26, 2008 & 640 & 20 & 0.6-m & I$_{\rm C}$ \\
   Aug 27, 2008 & 600 & 20 & 0.6-m & I$_{\rm C}$ \\ 
   Aug 28, 2008 & 570 & 20 & 0.6-m & I$_{\rm C}$ \\
   Oct 10, 2008 & 100 & ~\,1 & 1.6-m & H \\
   Sep 01, 2009 & 100 & ~\,1 & 1.6-m & H \\
\hline                                   
\end{tabular}
\end{table}

\section{Data reduction}

The reduction of the data was done with the usual \verb"IRAF" ~\verb"cl" ~tasks and consists of subtracting a master median bias image from each program image, and dividing the result by a normalized flat-field. In the H-band, additional steps of linearization and sky subtraction from dithered images were used in the preparation of the data. We treated the photometry extraction in the B- and I$_C$/H-bands in slightly different ways. The reason for this will become clear in the following discussion. 

\begin{figure}
 \resizebox{\hsize}{!}{\includegraphics{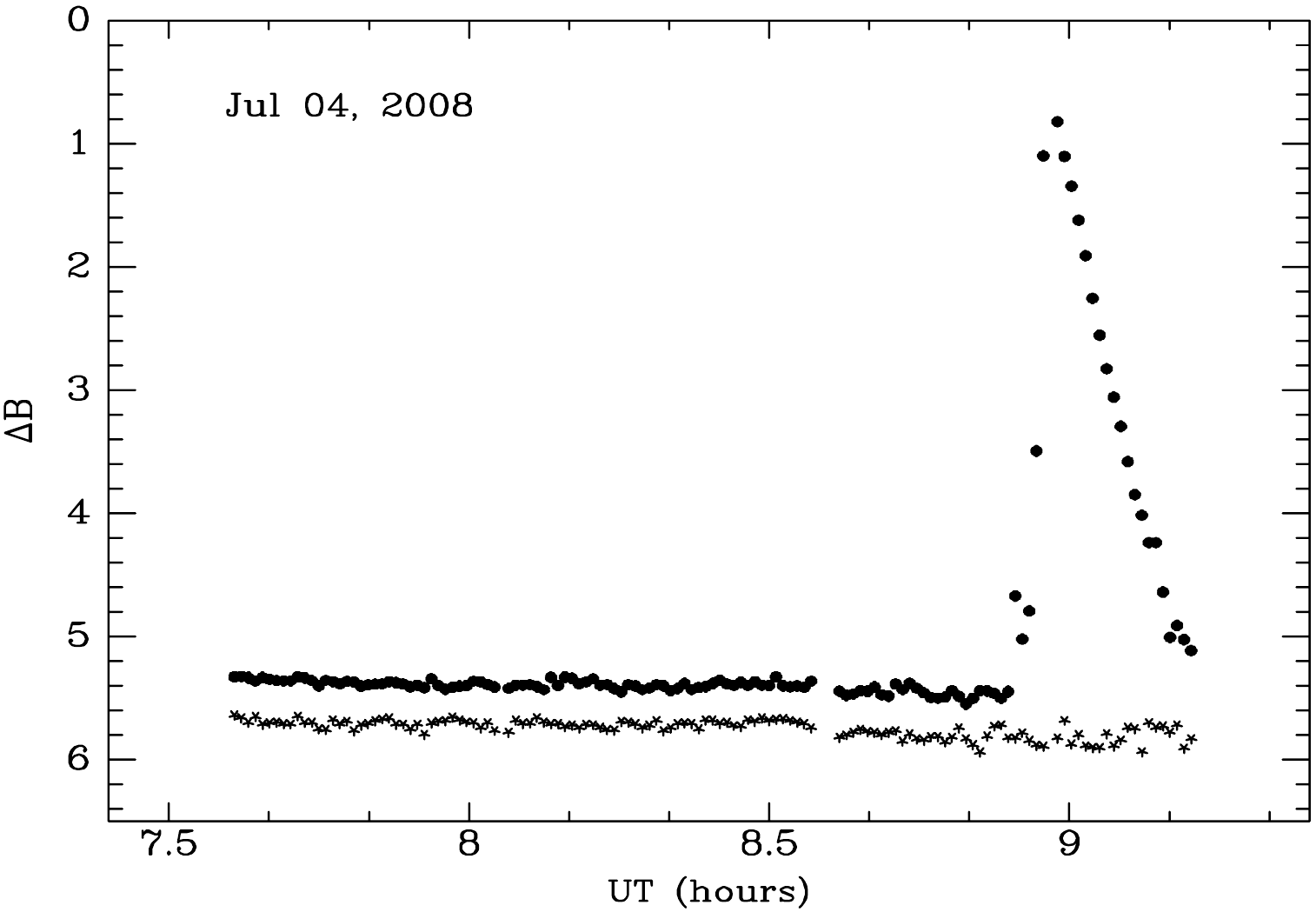}}
 \caption{B-band differential photometry of LHS~1070 on July 04, 2008. The light curves of LHS~1070 and of a comparison star are presented with full circles and stars, respectively.}
 \label{fig1}
\end{figure}

In the B-band, component  $\mathcal{A}$ dominates the flux of the system. This allows us to extract the relevant fluxes using plain aperture photometry. In the flare event (see Figure \ref{fig1}), since additional light could be coming from components $\mathcal{B}$ or $\mathcal{C}$, we fitted a double 2-D Moffat function to the stellar 
profile. 

In the I$_C$-band, since in principle both $\mathcal{B}$ and $\mathcal{C}$ components could contribute with substantial flux, we fitted simultaneously three 2-D Moffat functions to the stellar profile leaving only the position of component $\mathcal{A}$ and the amplitudes of the three components as free parameters to be searched for. 
Essentially the same procedure is used for the H-band data. The distances $\mathcal{A}$-$\mathcal{B}$ and $\mathcal{A}$-$\mathcal{C}$ were obtained from the orbital elements listed in Table~\ref{orbital_elements} for 
LHS~1070\,$\mathcal{A}$/($\mathcal{B}$+$\mathcal{C})$ and the orbital elements for LHS~1070\,$\mathcal{B}$/$\mathcal{C}$ from \citet{2008A&A...484..429S}. Our own measurements were used to improve the astrometric solution for the pair $\mathcal{A}$/($\mathcal{B}$+$\mathcal{C})$ (see Section \ref{improve_astrometry}). In all cases we used the \verb"amoeba"~routine of \citet{1992nrfa.book.....P} for the 
fitting procedure.

In order to increase the stability of the fits, for both bands we fixed the Moffat parameters derived  from the profile of star 6417-00147-1 in the Tycho catalog and used $\beta = 4.765$ as in \citet{2001MNRAS.328..977T}. This star is located at $\Delta\alpha=103.13\,{\rm arcsec}$ and $\Delta\delta=157.32\,\rm{arcsec}$ from LHS~1070\,$\mathcal{A}$.

\begin{figure}
 \resizebox{\hsize}{!}{\includegraphics{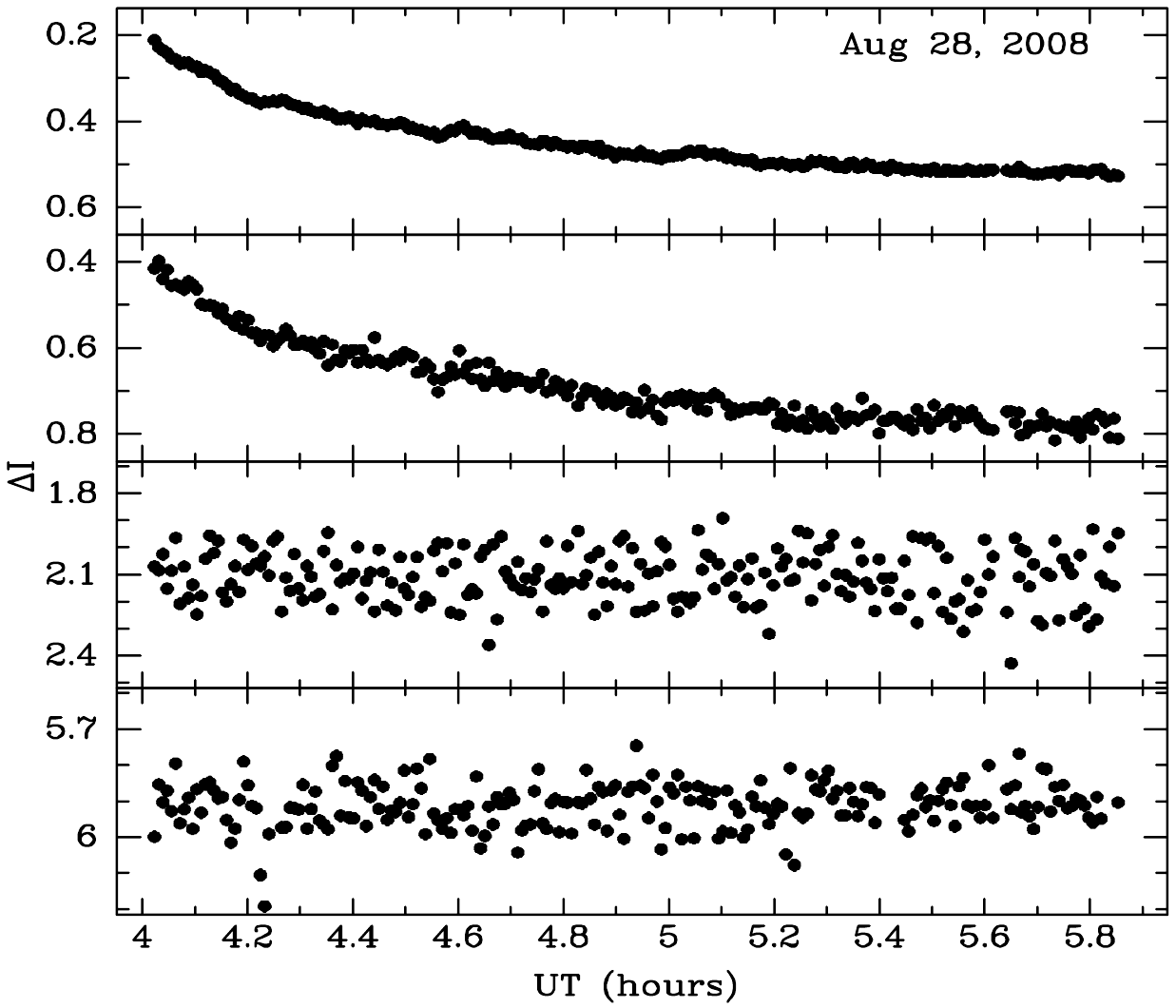}}
 \caption{I$_C$-band differential photometry of LHS~1070 on August 28, 2008. From top to bottom we show the aperture photometry for LHS~1070, 
the results of the 2-D Moffat function fitting for LHS 1070\,$\mathcal{A}$ and LHS~1070\,$\mathcal{B}$+$\mathcal{C}$ and the aperture photometry for a faint comparison star in the field.}
 \label{fig2}
\end{figure}

Figure~\ref{fig1} shows the B-band differential photometry for LHS~1070 on July 4, 2008. As one can see, a very strong flare characterized by an e-folding decay time of $\sim 165\,{\rm sec}$ and a factor of $\sim 80$ increase in brightness with respect to quiescence was observed. Figure~\ref{fig2} shows the I$_C$-band light curve of an event with a longer time-scale (the e-folding time is $\> \sim 1080\,{\rm sec}$) in August 28th. Unfortunately, the onset of this flare was not observed.

\section{Analysis and results}

\subsection{Astrometry}\label{se:astrometry}

\begin{figure}
\resizebox{\hsize}{!}{\includegraphics{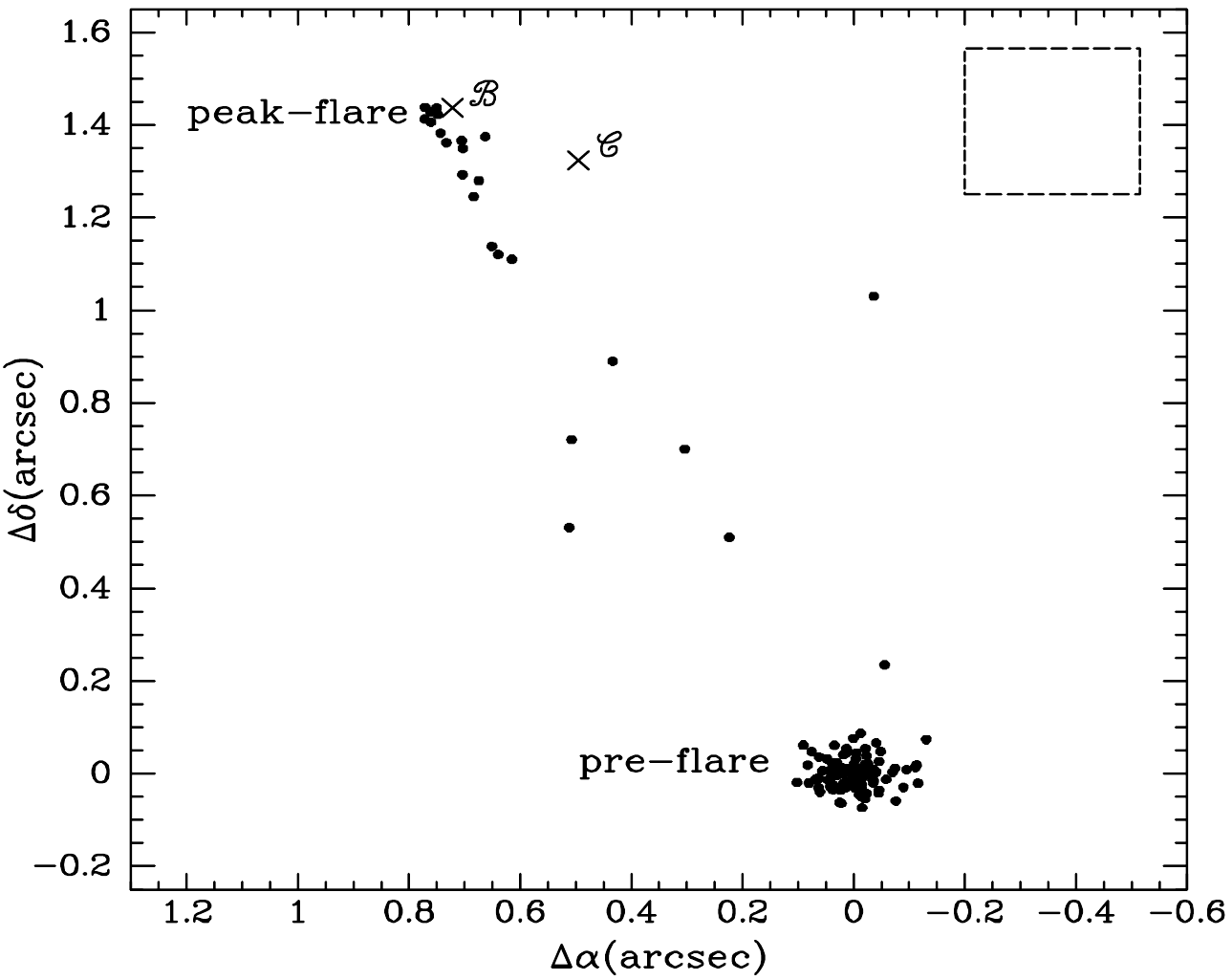}}
 \caption{The position of the photocenter of LHS~1070 with respect to a Tycho catalog star. The pre-flare position was obtained using the 101 images before the flare. Its uncertainty can be estimated from the bidimensional distribution of individual positions, and results in a positional accuracy of $\sim 20\,{\rm mas}$. The six images around the peak of the event define the position of the photocenter during the flare. The predicted positions of $\mathcal{B}$ and $\mathcal{C}$ components are marked with a ``$\times$"~symbol considering component $\mathcal{A}$ at the origin. The box shown in the upper right 
 corner represents the size of one pixel.}
 \label{fig3}
\end{figure}

\begin{figure}
\resizebox{\hsize}{!}{\includegraphics{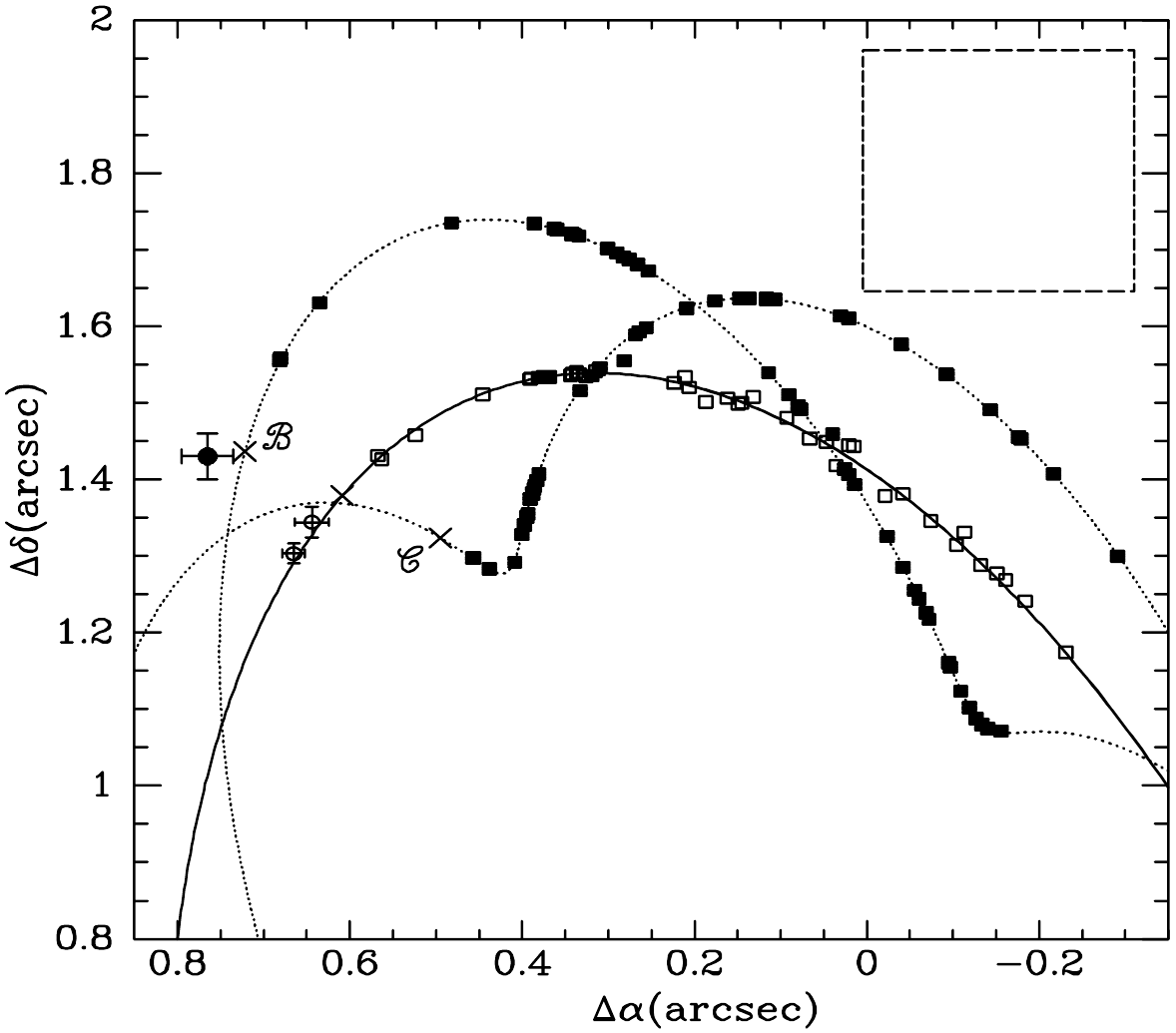}}
\caption{Configuration of LHS~1070 on Jul 04, 2008. The solid line represents the trajectory of the barycenter of LHS~1070\,$\mathcal{B}$+$\mathcal{C}$ around LHS~1070\,$\mathcal{A}$ which is at the origin. The dotted lines show the trajectories of components $\mathcal{B}$ and $\mathcal{C}$ around the barycenter of LHS~1070\,$\mathcal{B}$+$\mathcal{C}$, assuming the two components have the same mass. The positions measured by \citet{leinert2001} and \citet{2008A&A...484..429S} are shown as open and full squares. Our measurements appear as open circles. The full circle shows our estimated position for the flaring object. The predicted positions of the barycenter of LHS~1070\,$\mathcal{B}$+$\mathcal{C}$ and of components $\mathcal{B}$ and $\mathcal{C}$ are marked with a ``$\times$"\ symbol. The box shown in the upper right corner represents the size of one pixel of our detector}. 
\label{fig:orbita}
\end{figure}

Visual examination of the CCD images during the B-band flare suggests a significant displacement of the photocenter of LHS~1070 during the event. 101 pre-flare images allow the average relative position of LHS~1070\,$\mathcal{A}$ with respect to a reference star to be measured with $\sim20$ mas accuracy. As the flare progresses, the photocenter shifts toward North and East indicating that component $\mathcal{A}$ 
was not the flaring object. Figure \ref{fig3} shows the differential positions of LHS~1070 with respect to a reference object. To identify which component, $\mathcal{B}$ or $\mathcal{C}$, was responsible for the event, we proceeded as follows. First, we obtained an astrometric solution for the whole field using the registered pre-flare 
images. Seven stars (excluding LHS~1070) can be used for the astrometry. We used the \verb"IRAF" \verb"ccmap" ~task to obtain the plate scale (0.315 arcsec/pixel) and rotation of the images ($+0\fdg1$ with respect to North). The position corresponding to the flare was found by fitting a double 2-D Moffat function to the stellar profile in the six images closer to the flare peak. The flare position is shown together with 
the orbital solutions for LHS~1070\,$\mathcal{B}$/$\mathcal{C}$ \citep{2008A&A...484..429S} and LHS~1070\,$\mathcal{A}$/($\mathcal{B}$+$\mathcal{C}$) (Table \ref{orbital_elements}) in Figure \ref{fig:orbita}. We conclude that the flaring object was component $\mathcal{B}$.

\subsection{Astrometric solution and orbital fitting for LHS~1070 
$\mathcal{A}$/($\mathcal{B}$+$\mathcal{C})$}
\label{improve_astrometry}

\begin{figure}
 \resizebox{\hsize}{!}
 {\includegraphics[width=8.5cm,angle=-90]{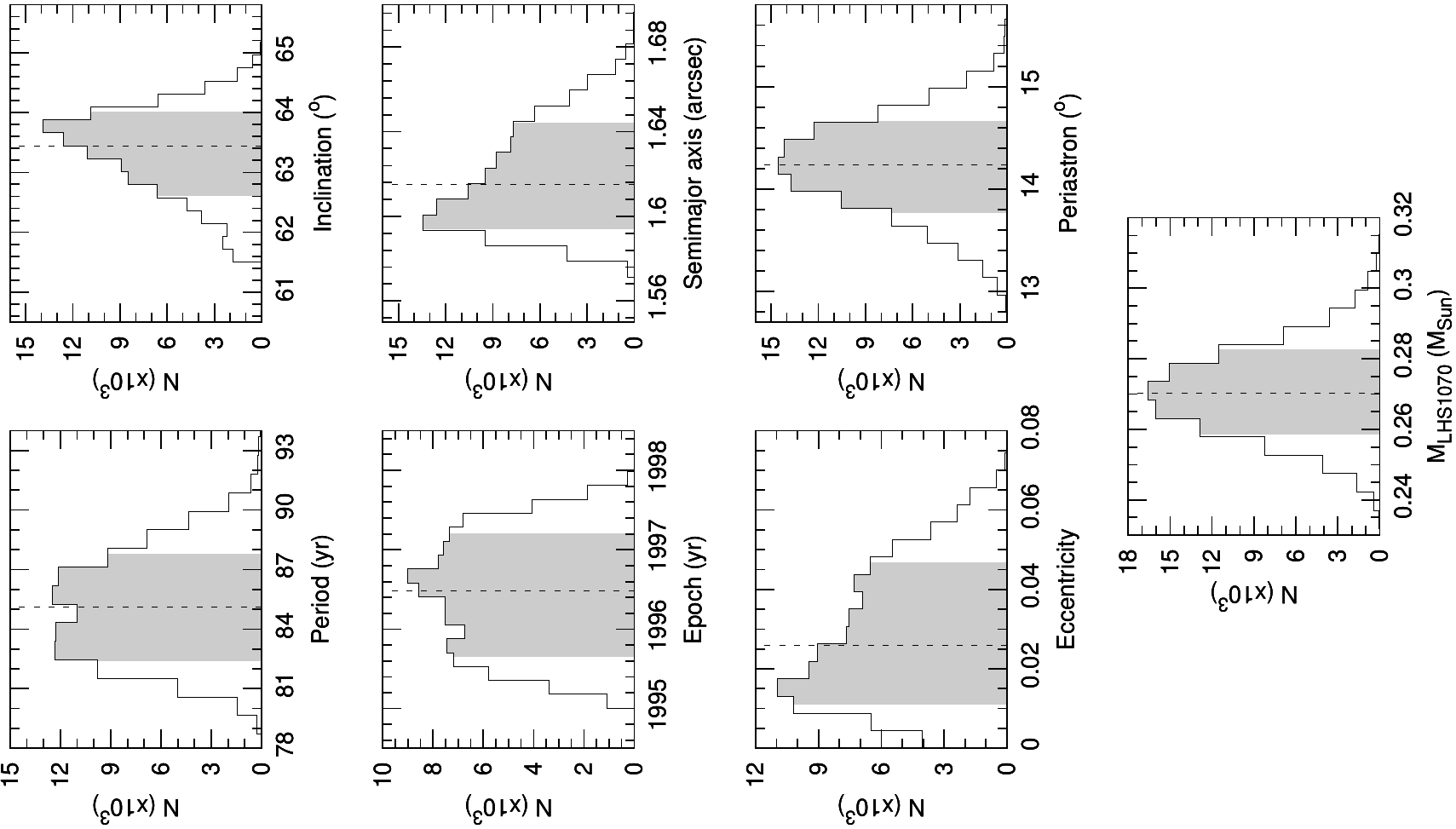}}
 \caption{Marginal distribution of posterior probability for the orbital elements 
          of LHS~1070\,$\mathcal{A}$/($\mathcal{B}$+$\mathcal{C}$). The dashed line 
          shows the median and the grey areas mark the $\pm 68\%$ confidence regions.}
 \label{histogram}
\end{figure}

We used the Monte Carlo Markov Chain approach to explore the distribution of probability of the orbital elements of LHS~1070\,$\mathcal{A}$/($\mathcal{B}$+$\mathcal{C}$). 
The data from \citet{2008A&A...484..429S}, \citet{leinert2001} and our own measurements (see Table \ref{our_mesure}) were used in this analysis. The orbital elements distributions of \citet{2008A&A...484..429S} were used as prior information to obtain the posterior distribution of the parameters. Figure~\ref{histogram} shows the marginal distributions and Table~\ref{orbital_elements} presents the numerical values with the associated $\pm 68\%$ uncertainties.

It is interesting to compare the results of the orbital elements obtained for the LHS~1070\,$\mathcal{A}$/($\mathcal{B}$+$\mathcal{C}$) system with the results obtained by \citet{2008A&A...484..429S} for LHS~1070\,$\mathcal{B}$/$\mathcal{C}$. The orbits of the three components are, within the uncertainties, coplanar.

\begin{table}
\caption{Distances between the photocenters of LHS~1070\,$\mathcal{B}$+$\mathcal{C}$ and 
         LHS~1070\,$\mathcal{A}$.}             
\label{our_mesure}      
\centering                          
\begin{tabular}{c c c }        
\hline\hline                 
Date&$\Delta\alpha$ ($^{\prime\prime}$)&  $\Delta\delta$ ($^{\prime\prime}$) \\    
\hline                        
Oct 10, 2008 & 0.644$\pm$0.022 & 1.344$\pm$0.022 \\      
Sep 01, 2009 & 0.665$\pm$0.013 & 1.303$\pm$0.013 \\
\hline                                   
\end{tabular}
\end{table}

\begin{table}
\caption{Orbital elements for the LHS~1070\,$\mathcal{A}$/($\mathcal{B}$+$\mathcal{C}$) binary.} 
\label{orbital_elements}      
\centering                          
\begin{tabular}{c c }        
\hline\hline                 
Parameter & Value   \\    
\hline                        
\vspace{2mm}
   P~(yr)     & 85.1$\pm2.7 $ \\
\vspace{2mm}
   $T_0$~(yr) & 1996.48$^{+0.72}_{-0.82}$ \\
\vspace{2mm}
   $a$~($''$) & 1.615$^{+0.029}_{-0.021}$ \\      
\vspace{2mm} 
   $e$        & 0.026$^{+0.021}_{-0.015}$    \\
\vspace{2mm}  
   $i$~($^{\circ}$) & 63.44$^{+0.57}_{-0.83} $     \\
\vspace{1.5mm} 
   $\Omega$~($^{\circ}$) & 14.24$^{+0.42}_{-0.47} $    \\

   M$_{\textrm{\tiny{LHS\,1070}}}~$(M$_\odot$) & 0.270$\pm0.012$    \\

\hline                                   
\end{tabular}
\end{table}

\subsection{The flare in LHS 1070 $\mathcal{B}$}

As one can see in Figure \ref{fig1}, the B-band flare is relatively fast, with a rise-time of $\sim$225$\,{\rm s}$. The decay back to quiescence took $\sim$880$\,{\rm s}$. The duration of the event is thus $\sim$1100$\,{\rm s}$. The flare amplitude with respect to the quiescent magnitude of component $\mathcal{A}$ is $\sim$4.8 mag. However, since component $\mathcal{B}$ is the flaring object, and adopting B $= 17.17$ and  B $= 20.57$ \citep{2000A&A...353..691L} as the quiescent B-magnitudes of components $\mathcal{A}$ and $\mathcal{B}$, we obtain $\sim$8.2 mag as a lower limit for the flare amplitude.

\subsubsection{Luminosity and total energy}

The luminosity of the flare can be estimated as follows. The monochromatic flux associated to the apparent quiescent magnitude B is,
\begin{equation}
 F_{\rm{\tiny{B}},\lambda} = F_{\tiny{\rm B}_0}10^{-\left(\frac{\tiny{\rm B}}{2.5}\right)},
\label{eq:fluxob}
\end{equation}
where $F_{\tiny{\rm B}_0} = 6.32\times10^{-9}\,\rm erg\,\rm cm^{-2}\,\rm s^{-1}$ $\mathring{\rm A}^{-1}$ is the absolute flux corresponding to zero magnitude \citep{1998A&A...333..231B}. The quiescent level observed prior to the flare is due to component $\mathcal{A}$. It corresponds to B=17.17 \citep{2000A&A...353..691L}. This sets a reference level to which the flare flux may be referred to. Thus, the quiescent level of component $\mathcal{A}$, expressed as luminosity in the B-band is,
\begin{equation}
  L_{\tiny{\rm B}} = 4\pi D^2F_{\tiny{\rm B}_0}\Delta\lambda 
  10^{-\left(\frac{\tiny{\rm B}}{2.5}\right)}\sim 5.4\times10^{27} \rm{erg}\,\rm{s}^{-1},
\label{eqlumib}
\end{equation}
where $\Delta\lambda = 890\,\mathring{\rm A}$ is the B-band FWHM \citep{2005ARA&A..43..293B} and $D=7.72\,\rm pc$ is the distance to the system. The luminosity of the event can be obtained using Eq. \ref{eqlumib} and the differential magnitudes of each point along the flare with respect to the quiescent level. The time-integral under the flare luminosity light curve is the total energy of the flare in the B-band. From this procedure we obtain $ E_{\rm \tiny B} \sim 9.8 \times 10^{31}$ erg.

\subsubsection{Magnetic field strength}

In order to obtain the magnetic field strength, we have to estimate the bolometric energy of the event, $E_{bol}$. We start with the flare energy in the B-band, $E_{\rm\tiny{B}}$, and use the relation among optical, UV and X-ray energies obtained from multispectral observations of flares in low-mass stars \citep{2005stam.book.....G}, 
\begin{equation}
E_{bol} \sim 20.9\,E_{\tiny{\rm B}} \sim 2.0\times10^{33}\,\rm ergs.
\label{eq:energiatotal}
\end{equation}

Knowing the total energy irradiated and assuming that this flare has the same mechanisms as in solar flares, we can estimate a lower limit for the magnetic field strength in the region of the event. Stellar flares are caused by sudden changes of magnetic field strength in the stellar corona as result of the reconnection of magnetic field lines \citep[see e.g.,][]{1976SoPh...50...85K, 1986SoPh..104....1P}. Such changes convert magnetic potential energy into plasma acceleration \citep[see e.g.,][]{1976SoPh...47..361K, 2007A&A...471..993D}. Part of the plasma is released from the star as coronal mass ejection and part is accelerated towards the chromosphere. The latter part interacts with denser plasma converting kinetic energy into thermal energy and is associated with optical and X-ray emission. According to \citet{1992ApJS...81..885H}, a blackbody at a temperature $\sim$ 10$^{4}\,\rm K$ can be used as a raw description of the optical emitting region. In terms of fractional area $X$ on the star 
\citep{2003ApJ...597..535H}, we have the flux at the maximum of the flare given by
\begin{equation}
 F_{\lambda} = X \frac{R^2_{\ast}}{D^2}\pi B_{\lambda}(T),
\end{equation}
where $T$ is the blackbody temperature, $B_{\lambda}$ is the Planck function, $R_{\ast}$ is the radius of the star and $D$ is the distance to the object. Using Eq. \ref{eq:fluxob} and $\Delta \rm B \sim 4.8$ for the maximum flare amplitude, the flux at the B-band pivotal wavelength is $F_{\rm B} \sim 7.12\times 10^{-14}\,\rm erg\,cm^{-2}\,s^{-1}\,\mathring{\rm A}^{-1}$. With the aid of the mass-radius relation for low-mass stars \citet{2009AIPC.1094..102C}, and $M=0.0815\,\rm M_{\odot}$ for LHS~1070\,$\mathcal{B}$ \citep{2000A&A...353..691L}, the fractional area obtained is $\sim 1.4\times10^{-2}$. The corresponding covered area and spherical volume are $\sim 1.4\times 10^{18}\rm\,cm^{2}$ and $\sim$1.7$\times10^{27}~{\rm cm}^3$, respectively.

Using the simplest case where the volume occupied by the magnetic field lines is spherical, and assuming the dimension of the acceleration region  equal to the emitting region \citep{2002paks.book.....A} and considering the standard model, i.e., the total flare energy being produced by magnetic energy decay \citep[see e.g.,][]{1994SoPh..153...19B, 2007A&A...471..993D}, we can estimate the magnetic field strength, $B$, using the relation:
\begin{equation}
 \frac{E_{bol}}{V} = \frac{B^2}{8\pi}.
\end{equation}

This gives a value of $B \sim 5.5$ kG in the flaring region. Notice that this is a lower limit. \citet{2007A&A...471L...5R} obtained an average $B$-field of $\sim 4$ kG over the entire star, from analysis of  spectroscopic data. We would expect to have larger values of the magnetic field strength in a flare.

\section{Discussion}

As one can see in Figure \ref{fig2}, component $\mathcal{A}$ was responsible for the flare on August 28, 2008. The photometric activity detected in both $\mathcal{A}$ and $\mathcal{B}$ components is consistent with the spectroscopic results of \citet{2000A&A...353..691L}.

Photometric activity was unknown in LHS~1070 before. The noticeable flare in LHS~1070\,$\mathcal{B}$ shows that even close to the Hydrogen-burning limit, relatively old objects can still show significant magnetic fields. The activity level observed in the components of the LHS~1070 system, and the age estimated by \citet{2007A&A...471L...5R} are in agreement with the activity lifetime-spectral type relation discussed by \citet{2008AJ....135..785W}. 

In order to stress the importance of the event observed in LHS 1070\,$\mathcal{B}$, we recollect the most impulsive optical flares previously recorded on dMe stars. Impulsive stellar flares are events lasting $\sim$100-1000\,s with large amplitudes. \citet{2005stam.book.....G} describes several flares among which the most impulsive was the one observed by de \citet{1989A&A...211..157D} on a dM5.5e star, UV Cet, on December 23, 1985, with amplitude of 5 mag in the B-band. Another impulsive flare with 7 mag amplitude at blue wavelengths was observed by \citet{bond1976} on a dMe star. \citet{2006A&A...460L..35S} reported a multi-wavelength observation of a flare on LP 412-31. This object has spectral type M8 and showed a 6 mag amplitude flare in the V-band. More recently, \citet{2010ApJ...714L..98K} observed a flare on the dM4.5e star YZ CMi with amplitude $\sim 6$ mag in the U-band. Thus, since the lower limit for the LHS~1070\,$\mathcal{B}$ flare presented here is $\sim$8.2 mag in the B-band, we conclude that this event has the largest amplitude ever observed in a flare star. 

Regarding the energy released in flares, \citet{2005stam.book.....G} discusses a few events observed in BY Dra and AD Leo in the B-band that reached $\sim 10^{35}$ erg. However, the time-scales for these events are larger than the observed by us. Besides, those objects have spectral types earlier than that of LHS 1070 $\mathcal{B}$. This means that a fair comparison between the energy released in the event observed on LHS 1070 $\mathcal{B}$ and other objects should be restricted to similar spectral types and similar time-scales. Using these criteria there is only one object that has a similar energy budget in a flare: the M8 dwarf, LP 412-31 for which $E_{bol} \sim 6\times10^{32}$ erg.

\begin{acknowledgements}
      We thank Dr. Joaquim E. R. Costa and Dr. Carlos Alberto P. C. O. Torres for helpful 
      suggestions. This work was supported by Coordena\c{c}\~ao de Aperfei\c{c}oamento 
      de Pessoal de N\'ivel Superior (CAPES). This research is based on observations 
      carried out with facilities of Laborat\'orio Nacional de Astrof\'isica (LNA/MCT) 
      in Brazil.
\end{acknowledgements}


\begin{thebibliography}{}

\bibitem[Aschwanden(2002)]{2002paks.book.....A} Aschwanden, M. J. 2002, Space Sci. Rev., 101, 1

\bibitem[Berger(2006)]{2006ApJ...648..629B} Berger, E.\ 2006, \apj, 648, 
629 

\bibitem[Bessell et al.(1998)]{1998A&A...333..231B} Bessell, M.~S., Castelli, F., \& Plez, B.\ 1998, \aap, 333, 231 

\bibitem[Bessell(2005)]{2005ARA&A..43..293B} Bessell, M.~S.\ 2005, \araa, 43, 293 

\bibitem[Bond(1976)]{bond1976} Bond, H. E. 1976, Inf. Bull. Variable Stars, 1160

\bibitem[Brown et al.(1994)]{1994SoPh..153...19B} Brown, J.~C., et al.\ 
1994, \solphys, 153, 19 

\bibitem[Chabrier et al.(2009)]{2009AIPC.1094..102C} Chabrier, G., Baraffe, 
I., Leconte, J., Gallardo, J., \& Barman, T.\ 2009, American Institute of Physics Conference Series, 1094, 102 

\bibitem[Costa et al.(2005)]{2005AJ....130..337C} Costa, E., M{\'e}ndez, 
R.~A., Jao, W.-C., Henry, T.~J., Subasavage, J.~P., Brown, M.~A., Ianna, 
P.~A., \& Bartlett, J.\ 2005, \aj, 130, 337 

\bibitem[Dauphin(2007)]{2007A&A...471..993D} Dauphin, C.\ 2007, \aap, 471, 993 

\bibitem[de Jager et al.(1989)]{1989A&A...211..157D} de Jager, C., et al.\ 1989, \aap, 211, 157 

\bibitem[Gershberg(2005)]{2005stam.book.....G} Gershberg, R.~E.\ 2005, 
Solar-Type Activity in Main-Sequence Stars, ed. R. E. Gershberg

\bibitem[Hawley \& Fisher(1992)]{1992ApJS...81..885H} Hawley, S.~L., \& Fisher, G.~H.\ 1992, \apjs, 81, 885 

\bibitem[Hawley et al.(2003)]{2003ApJ...597..535H} Hawley, S.~L., et al.\ 2003, \apj, 597, 535 

\bibitem[Kopp \& Pneuman(1976)]{1976SoPh...50...85K} Kopp, R.~A., \& Pneuman, G.~W.\ 1976, \solphys, 50, 85 

\bibitem[Kowalski et al.(2010)]{2010ApJ...714L..98K} Kowalski, A.~F., 
Hawley, S.~L., Holtzman, J.~A., Wisniewski, J.~P., 
\& Hilton, E.~J.\ 2010, \apjl, 714, L98 

\bibitem[Kuperus(1976)]{1976SoPh...47..361K} Kuperus, M.\ 1976, \solphys, 47, 361

\bibitem[Leinert et al.(2000)]{2000A&A...353..691L} Leinert, C., Allard, F., Richichi, A., \& Hauschildt, P.~H.\ 2000, \aap, 353, 691

\bibitem[Leinert et al.(2001)]{leinert2001} Leinert, C., Jahrei{\ss}, H., Woitas, J., Zucker, S., Mazeh, T., Eckart, A., {\ K\&ouml}hler, R.\ 2001, \aap, 367, 183 

\bibitem[Press et al.(1992)]{1992nrfa.book.....P} Press, W. H., Teukolsky, S. A., Vetterling, W. T., \& Flannery, B. P. 1992, Numerical Recipes in Fortran: The Art of Scientific Computing (Cambridge: Cambridge Univ. Press)

\bibitem[Priest(1986)]{1986SoPh..104....1P} Priest, E.~R.\ 1986, \solphys, 104, 1 

\bibitem[Reiners et al.(2007)]{2007A&A...471L...5R} Reiners, A., Seifahrt, A., K{\"a}ufl, H.~U., Siebenmorgen, R., \& Smette, A.\ 2007, \aap, 471, L5 

\bibitem[Skumanich(1972)]{1972ApJ...171..565S} Skumanich, A.\ 1972, \apj, 
171, 565 

\bibitem[Seifahrt et al.(2008)]{2008A&A...484..429S} Seifahrt, A., R{\"o}ll, T., Neuh{\"a}user, R., Reiners, A., Kerber, F., K{\"a}ufl, H.~U., Siebenmorgen, R., \& Smette, A.\ 2008, \aap, 484, 429 

\bibitem[Stelzer et al.(2006)]{2006A&A...460L..35S} Stelzer, B., Schmitt, J.~H.~M.~M., Micela, G., \& Liefke, C.\ 2006, \aap, 460, L35 

\bibitem[Trujillo et al.(2001)]{2001MNRAS.328..977T} Trujillo, I., Aguerri, J.~A.~L., Cepa, J., \& Guti{\'e}rrez, C.~M.\ 2001, \mnras, 328, 977 

\bibitem[West et al.(2008)]{2008AJ....135..785W} West, A.~A., Hawley, 
S.~L., Bochanski, J.~J., Covey, K.~R., Reid, I.~N., Dhital, S., Hilton, 
E.~J., \& Masuda, M.\ 2008, \aj, 135, 785 

\end{thebibliography}
\end{document}